\documentclass{JHEP3}

\usepackage{graphicx,psfrag,verbatim}
\usepackage{latexsym}
\usepackage{amsmath}
\usepackage{amsfonts}
\usepackage{amssymb}
\def\look
{\framebox{$\clubsuit$}\,}
\newcommand{\be}{\begin{equation}}
\newcommand{\ee}{\end{equation}}
\newcommand{\bea}{\begin{eqnarray}}
\newcommand{\eea}{\end{eqnarray}}

\newcommand{\ben}{\begin{eqnarray}}
\newcommand{\een}{\end{eqnarray}}
\newcommand{\bes}{\begin{subequations}}
\newcommand{\ees}{\end{subequations}}

\title{Holographic Description of Heavy-Quark Potentials in an Inflationary Braneworld Scenario}

\author{Luciano Barosi $^{a}$, Francisco A. Brito $^{a}$ and Amilcar R. Queiroz $^{b,c,d}$ \\
$^{a}$Departamento de F\'\i sica, Universidade Federal de Campina
Grande, Caixa Postal 10071, 58109-970  Campina Grande, Para\'{\i}ba,
Brazil\\
$^{b}$International Center for Condensed Matter Physics (ICCMP),
Universidade de Bras\'\i lia, Caixa Postal 04667, Bras\'\i lia, DF,
Brazil\\
$^c$  Universidade Federal de Goi\'as, Campus Catal\~ao, Departamento de F\'{\i}sica, St. Universit\'ario - 75700-000, Catal\~ao-GO, Brazil \\
$^d$ Instituto de F\'{\i}sica, Universidade de Bras\'{\i}lia, Caixa Postal: 04455, CEP.: 70919-970 - Bras\'{\i}lia, Brazil

\\
E-mail:{ lbarosi@ufcg.edu.br, fabrito@df.ufcg.edu.br,
amilcarq@gmail.com}}

\abstract{We study heavy-quark potential in an inflationary braneworld scenario. The scenario we consider is an (Euclidean) $AdS_4$ inflating braneworld embedded in a static Euclidean $AdS_5$. Using gauge/gravity duality we obtain a confining phase depending on the ratio between the Hubble parameter $H$ in the braneworld and the brane tension $\sigma$.}

\keywords{Holographic QCD, Quark-antiquark Potential, Brane Cosmology}

\begin{document}
%\maketitle

%%%%%%%%%%%%%%%%%%%%%%%%%%%%%%%%%%%
\section{Introduction}
%%%%%%%%%%%%%%%%%%%%%%%%%%%%%%%%%%

The holographic computation of heavy quark potentials
\cite{Maldacena:1998im,Rey:1998ik,Minahan:1998xq,Greensite:1999jw,Sonnenschein:2000qm,BoschiFilho:2004ci,BoschiFilho:2005mw,Andreev:2006ct} is formulated by relating the
string world-sheet area along the bulk with the area of the Wilson loop
whose contour lies on the $p$-dimensional boundary of the bulk.  The bulk
is described by the geometry of a stack of a large number $N$ of
$Dp$-branes in ten-dimensional spacetime. Following the usual Euclidean
lattice QCD\footnote{For a recent and comprehensive review on QCD from the AdS/CFT point of view see \cite{Erdmenger:2007cm}. For an interesting criticism on this point of view see \cite{Csaki:2008dt}.} formulation for the confining regime, one usually assumes the
Euclidean signature to evaluate both the Wilson loop and the
string world-sheet area \cite{Maldacena:1998im,Sonnenschein:2000qm,Witten:1998zw,Polyakov:2000ti,Bigazzi:2004ze},
although
similar evaluations in Lorentzian signature are also possible
\cite{Drukker:1999zq}. For AdS/CFT correspondence \cite{Maldacena:1998im}
one finds a
Coulomb-like potential. However, in order to find a gravity dual for a
confining theory, such as QCD, one should `deform' the $AdS$ geometry of
the bulk
\cite{BoschiFilho:2004ci,BoschiFilho:2005mw,Csaki:2008dt,Klebanov:2000hb,Maldacena:2000yy,BoschiFilho:2006pe}.
The deformation we assume here is a deviation from the conformal case that
is {\it naturally} achieved in geometries of brane cosmology scenarios
\cite{Binetruy:1999ut, Binetruy:1999hy}. In order to keep our setup in the string theory framework, this deformation is such that almost all of the well-known AdS/CFT correspondence is maintained. As we shall see later, this is a slight deviation around AdS space controlled by the tension and the cosmological constant on the brane. We have found that such a
deviation implies a {\it confining regime} in the dual gauge field theory
that is directly associated with the cosmological constant on the brane.
Indeed, this is a general class of geometries where the brane has a
cosmological constant, i.e., a {\it bent} brane, such as $AdS_4$ or $dS_4$
branes
\cite{Cvetic:1993xe,Kaloper:1999sm,Karch:2000ct,Bazeia:2006ef,Bazeia:2007vx,Bazeia:2004yw,Ghoroku:2006nh},
in contrast with flat branes in the original AdS/CFT correspondence.

In this paper we shall assume an inflationary phase on the brane \cite{Binetruy:1999ut,Binetruy:1999hy,Bazeia:2007vx,Dvali:1998pa,Lukas:1998yy,Chamblin:1999ya,Lukas:1999yn}. In
order to compute a Wilson loop on an inflationary brane we euclideanize
the time --- for further details on Euclidean brane cosmology see
\cite{Savonije:2001nd,Medved:2001jq}. For consistency, the inflationary
scenario is equivalent to assume an `Euclidean $AdS_4$' brane with
negative cosmological constant $\bar{\Lambda}$. We shall use the
well-known brane cosmology metric \cite{Binetruy:1999ut,Binetruy:1999hy}. In this case the Nambu-Goto action and the interquark distance scale
differently with time. We show that strings live in a
static background in the bulk, whereas their endpoints describe a gauge
field theory on the boundary with an Euclidean $AdS_4$ metric. Thus, in order to
compute the static quark potential on the boundary, the Wilson loop is
evaluated in the ``comoving frame".

The main result we obtain in this work is an explicit relation between interquark potential and inflation. We show explicitly that the interquark potential in the confining phase depends on the Hubble parameter and on the tension of the braneworld. The Hubble parameter here is taken as a constant. This is a signature of an inflationary phase in the braneworld. Furthermore, in this inflationary scenario the Hubble parameter can be seen as an order parameter. When the Hubble parameter is zero the interquark potential is Coulomb-like, indicating a deconfining phase. As the Hubble parameter is turned on, the confining phase is turned on.

In the present work we shall regard a delta function source for a single inflating
3-brane at $r\!=\!0$ with tension $\sigma$ as the place where there are
$N$ coincident `deformed $D3$-branes' generating the same background \cite{Bazeia:2007vx}. We
identify the $D3$-brane tension $T_{D3}=\sigma$. At the boundary of this
spacetime generated by such deformed $D3$-branes, we have a
four-dimensional Euclidean $AdS_4$ geometry where the Wilson loop is
evaluated and so does the heavy quark-anti-quark potential associated with
the dual gauge theory.

This paper is organised as follows: in section 2, we describe the brane cosmology scenario we are going to work with; in section 3, we consider an inflating braneworld embedded in a static $AdS_5$ and then obtain the interquark distance integral and the action integral;
in section 4, we set the large $N$ limit (or large $D$-brane tension) and describe the cases in which the interquark distance and the action should be evaluated; in section 5, we obtain the interquark potential in the cases delimited in the previous section. In particular, we explicit show the Hubble parameter dependence of the interquark potential in the confining phase. We close this paper with some conclusions and perspectives.

%%%%%%%%%%%%%%%%%%%%%%%%%%%%%%%%%%%%%%%%%%%%
\section{Brane Cosmology Background}
%%%%%%%%%%%%%%%%%%%%%%%%%%%%%%%%%%%%%%%%%%%%

The metric due to a stack of a large number $N$ of $D3$-branes provides a nice gravity dual for a gauge field theory developing both confining and non-confining regimes. This background is a ten-dimensional spacetime with metric
\ben ds^2=\left(1+\frac{R^4}{U^4}\right)^{-1/2}( -dt^2+ d\vec{x}^2 )+\left(1+\frac{R^4}{U^4}\right)^{1/2}(dU^2+U^2d\Omega_5^2).\een
Sufficiently close to the brane, i.e., $U\ll R$ the space has the $AdS_5\times S^5$ 
topology. As usual $R$ is the $AdS_5$ radius (the same as the $S^5$ radius) and $d\Omega_5$ is the $S^5$ metric. Disregarding the $S^5$ part, we are left with an $AdS_5$ spacetime with metric 
\ben
\label{D3brane} 
ds^2=\alpha'\left[\frac{U^2}{R^2}( -dt^2+ d\vec{x}^2 )+\frac{R^2}{U^2}dU^2\right],
\een
rescaled in terms of string units, whose boundary at $U\to\infty$ is a four-dimen\-sion\-al Min\-kowski 
space. One may euclideanize the boundary of the spacetime in order to calculate a Wilson loop in this boundary. The Wilson loop area is related to the string world-sheet area. The five-dimensional spacetime in this case is considered as an Euclidean $AdS_5$ space \cite{Savonije:2001nd,Medved:2001jq}.

%%%%%%%%%%%%%%%%%%%%%%%%%%%%%

Similar spaces are obtained in braneworld scenarios
with inflation \cite{Binetruy:1999ut, Binetruy:1999hy, Bazeia:2007vx} or without inflation \cite{Randall:1999vf} --- see \cite{BoschiFilho:2005mw} for developments on gauge/gravity duality in the Randall-Sundrum scenario \cite{Randall:1999vf}. Braneworld scenarios are found by considering a delta function source in a five-dimensional gravity theory given by the action \ben\label{action}S\!=\!-\frac{1}{2\kappa_{5}^2}\!\!\int{\!d^5x\sqrt{-g}(R+\Lambda_{bulk})
+\sigma\int{\!d^5x\sqrt{-g}\,\delta(r)}}, \een
where it is assumed that 
$\kappa_4^2=8\pi G_N\simeq{\kappa_5^4\sigma}/{6}=2$, with $G_N$ being Newton's constant and $\sigma$ is going to be associated with the brane tension. The five-dimensional cosmological constant is defined as
$\rho_{bulk}\equiv \Lambda_{bulk}=-\sigma^2\kappa_5^2/6$, which
satisfies the Randall-Sundrum fine tuning \cite{Randall:1999vf}. For a brane cosmology with an inflating 3-brane \cite{Binetruy:1999ut, Binetruy:1999hy, Bazeia:2007vx} the spacetime general metric is of the form \ben
\label{a_y_r}
ds^2=-n^2(t,r)dt^2+a^2(t,r)\gamma_{ij}dx^idx^j+b^2(t,r)dr^2 ,
\een 
where $\gamma^{ij}$ is a maximally symmetric 3-dimensional metric
with spatial curvature $k=-1,0,1$. The cosmological solution for a 3-brane inflating along $t,x^i$, but static in $r$ (i.e., along the bulk) \cite{Binetruy:1999ut, Binetruy:1999hy} reads 
\ben\label{a_y_r2}
a(t,r)\!&=&\!a_0(t)
\left[\frac{1}{2}\left(1+\frac{\kappa_5^2\rho_b^2}{6\rho_{bulk}}\right)
+\frac{1}{2}\left(1-\frac{\kappa_5^2\rho_b^2}{6\rho_{bulk}}\right)\cosh(\mu
r)\right.\left.-\frac{\kappa_5\rho_b}{\sqrt{-6\rho_{bulk}}}\sinh(\mu
|r|)\right]^{1/2}\!\!\!\!\!,\!\! 
\nonumber \\  n(t,r)\!&=&\!\frac{\dot{a}(t,r)}{\dot{a}_0(t)} \nonumber \\ b(t,r)\!&=&\!1, \een
where
$\mu=(1/3)(\kappa_5^4\sigma^2)^{1/2}$, $a_0$ is the scale factor on the brane and $\rho_b$ is the brane energy density. The
radiation `${\cal C}-$term' and the curvature `$k-$term' have been disregarded. The dynamics on the 3-brane  \cite{Bazeia:2007vx} is governed by the induced Friedmann equation on the brane 
\ben\label{friedman}H^2=\frac{2}{3}\rho\left(1+\frac{\rho}{2\sigma}\right),
\een
where $H={\dot{a}_0}/{a_0}$, with $a_0\!=\!a(t,r\!=\!0)$ being the scale factor on the brane worldvolume
with metric
\ben \label{metric4d}
ds_4^2=-dt^2+a_0^2(t)(dx^2+dy^2+dz^2),
\een
where we have assumed that $\rho_b=\rho+\sigma$, with $\sigma$ being the brane tension and $\rho$ being a cosmological energy in the brane.

For $r\gg R\sim 1/\mu$, with $R$ being a characteristic radius of the bulk, one may write $a(t,r)=U(r)a_0(t)\sim e^{r/R}a_0(t)$. If $U/R\ll1$ then $dU=(U/R)dr\sim 0$, and the metric (\ref{a_y_r}) can be reduced to a metric 
similar to that of a space due to a large number $N$ of $D3$-branes (\ref{D3brane}), except for the fact that its boundary at $U\to\infty$ is {\it not} a flat spacetime. 
We anticipate that the original metric (\ref{a_y_r}) with the solutions (\ref{a_y_r2}) may be written in the general form\footnote{There is a suitable rescaling of $U$ and $R$ so that this metric can be compared with (\ref{D3brane}). For further details see ahead the transformations (\ref{0}).}
\begin{equation}\label{mod_defayet_metric}
	ds^2=\alpha'\left[\frac{U^2}{R^2}\left( -dt^2+a_0(t)^2 d\vec{x}^2 \right) + \frac{R^2 U^2}{\left(U^2-C_1 \right)\left(U^2-C_2 \right)}~dU^2\right],
\end{equation}
with $C_{1,2}$ being proportional to the brane properties, for instance its tension $\sigma$ and its geometry  --- see below for further details. 
Particularly in the inflationary case, the space has a positive cosmological constant, i.e., a four-dimensional de Sitter space, with Lorentzian signature. 
In the boundary, the Wilson loop is calculated in the Euclideanized space, that means an `Euclidean $AdS_4$' space. The five-dimensional spacetime in this case is considered as an `Euclidean $AdS_4$-$AdS_5$' space \cite{Savonije:2001nd,Medved:2001jq}.

%%%%%%%%%%%%%%%%%%%%%%%%%%%%%%%%%%%%%%%%%%%%%%%%%%%%%%%%%%%%%%%%%%%%%%%%
\section{The gauge/gravity duality for an inflating brane background}
%%%%%%%%%%%%%%%%%%%%%%%%%%%%%%%%%%%%%%%%%%%%%%%%%%%%%%%%%%%%%%%%%%%%%%%%

In this section we analyse the case in which $H$, the Hubble parameter for the braneworld is a constant. This is a signature of an inflationary phase in the braneworld. 

For future reference, we begin by rewriting the metric (\ref{a_y_r}) in the general form
\begin{equation}
	ds^2=n^2(t,r)~edt^2+a^2(t,r)\gamma_{ij}dx^idx^j+b^2(t,r)dr^2,
\end{equation}
where $e=-1$ means Lorentzian signature and $e=+1$ (i.e. $t\mapsto i\tau)$ means Euclidean signature. In this case we may rewrite the scale factor in (\ref{a_y_r2}) as
\begin{equation}
	a(t,r)=U(r)a_0(t),
\end{equation}
where the ``warp factor" reads
\begin{equation}
	U(r)=\left(e\gamma+(1-e\gamma)\cosh(\mu r)-\sqrt{1-2e\gamma}\sinh(\mu |r|)\right)^{1/2},
\end{equation}
with 
\begin{eqnarray}
	\gamma&=&\frac{3H^2}{2\sigma}~~ \textrm{ when } e=-1~~\textrm{or }\:\:\ \\ \gamma&=&\frac{3H_E^2}{2\sigma}~~\textrm{ when } e=+1,
\end{eqnarray}
with $H_E=\frac{\dot{a_0}(\tau)}{a_0(\tau)}$, where the subindex $E$ stands for Euclidean, and $\sigma$ being the 3-brane tension. Observe that $ H\to iH_E$ or $\gamma\to -\gamma$ when $e=-1\mapsto e=+1$. 

In the Euclidean case ($e=+1$), we mostly consider this geometry as an asymptotically ($\mu r\to\infty$) Euclidean $AdS_4$-$AdS_5$ space. We also consider the Euclidean Hubble parameter $H_E^2=\frac{2}{3}\gamma\sigma=\textrm{const.}$. This implies that $a_0(\tau)=\exp{(\pm \sqrt{\bar{\Lambda}}\tau)}$, where
$\bar{\Lambda}=\frac{2}{3}\gamma\sigma$ is defined as the cosmological constant on the Euclidean brane. As we   observe later, in order to ensure leading string contributions, the brane tension should be very large. Thus we are going to analyse the case when $\sigma\to \infty$. In this case, we may have three possible regimes for $\gamma$ depending on $H_E$, i.e.,
\begin{enumerate}
	\item $\gamma\to 0$, when $H_E$ is held fixed,
	\item $\gamma$ is finite, when $H_E^2$ goes to infinity as fast as $\sigma$ goes to infinity, and
	\item $\gamma\to \infty$, when $H_E^2$ goes to infinity faster than $\sigma$ goes to infinity.
\end{enumerate}
Later we will see how $\gamma$ is related to a confining phase for the interquark potential.
In order to write the interquark distance and the action in terms of the string world-sheet we follow \cite{Sonnenschein:2000qm} and define
\begin{eqnarray}\label{ff}
	f(\tau,r)&=&\sqrt{g_{\tau \tau}~g_{xx}}=U(r)^2a_0(\tau)\equiv f(r)a_0(\tau), \\
	g(\tau,r)&=&\sqrt{g_{\tau \tau}~g_{rr}}=U(r)\equiv g(r),\label{fg}
\end{eqnarray}
where the subindices $\tau$, $x$ and $r$ in the metric $g_{\mu \nu}$ stand for the (Euclidean) time coordinate, the coordinate along the brane and the coordinate tranverse to the brane, respectively.
Now, the dependence of the interquark distance $L$ with respect to $a_0(\tau)$ is
\begin{equation}\label{Length}
	L(\tau)=2\int_{r_0}^{r_1} dr ~ \frac{g(\tau,r)}{f(\tau,r)}\frac{f(\tau,r_0)}{\sqrt{f(\tau,r)^2-f(\tau,r_0)^2}}=L_0~\frac{1}{a_0(\tau)},
\end{equation}
where $L_0$ is the static length given by
\begin{eqnarray}\label{stat_len}
	L_0&=&2\int_{r_0}^{r_1} dr ~ \frac{g(r)}{f(r)}\frac{f(r_0)}{\sqrt{f(r)^2-f(r_0)^2}},\nonumber \\ 
           &=&2\int_{r_0}^{r_1} dr ~ \frac{1}{U(r)}~\frac{U(r_0)^2}{\sqrt{U(r)^4-U(r_0)^4}}.
\end{eqnarray}
The limits $r_0$ fixed and $r_1\to\infty$
stand for the place where $U(r)$ develops a minimum and the boundary of the five-dimensional spacetime,
respectively. As we   see shortly, despite of the dependence of the length (\ref{Length}) on the scale $a_0(\tau)=\exp{(\pm \sqrt{\bar{\Lambda}}\tau)}$, the area of the string world-sheet is invariant under
such a scaling. This clearly generates a conflict. Roughly speaking, this happens because the area in the expectation value of the Wilson loop 
\ben 
<\!W({\cal C})\!\!>\,\sim e^{-const. L_0\int_0^T{d\tau/a_0(\tau)}}, 
\een 
diverges exponentially, whereas the string world-sheet area diverges linearly as $T\to\infty$. Thus, in order to circumvent this problem we keep the calculations of the Wilson loop in a flat spacetime at the boundary $r=r_1$, by assuming that the Wilson loop is being calculated in a ``comoving frame". This is the frame with static length $L_0=a_0(\tau)L(\tau)$ given in (\ref{stat_len}). Therefore one ensures an area law for 
\begin{equation}
\label{Cont}
<\!W({\cal C})\!\!>\,\sim e^{-const. L_0T}, 
\end{equation} 
that gives a confining phase. Alternatively, one
could also consider calculations by imposing a stringent bound on $\bar{\Lambda}$, i.e, $\sqrt{\Lambda}T\ll1$, such that we could use $L(\tau_0)\simeq L_0$. However, this is precisely the limit of a flat four-dimensional space on the 3-brane, where $\gamma\simeq 0$, or simply
the $AdS_5$ limit which is a conformal space and is not a good gravity dual for a gauge field theory developing a confining regime.

As we have anticipated, the string world-sheet action (i.e. the Nambu-Goto action) is not dependent on the scale $a_0(\tau)$. After few calculations, one can show that it is written as 
\begin{eqnarray}\label{action0}
	S&=& \frac{1}{2\pi \alpha'}\int_{r_0}^{r_1} \int_{0}^{T}dr d\tau\frac{g(r,\tau)f(r,\tau)}{\sqrt{f(r,\tau)^2-f(r_0,\tau)^2}},\nonumber   \\ 
         &=& \frac{T}{2\pi \alpha'}\int_{r_0}^{r_1} dr \frac{U(r)^3}{\sqrt{U(r)^4-U(r_0)^4}},
\end{eqnarray}
Observe that the dependence of the integrand with respect to $\tau$ has disappeared. The area enclosed by some square contour ${\cal C}$ in (\ref{Cont})
can be scaled up by $a_0(\tau)$, whereas the world-sheet area in (\ref{action0}) cannot be scaled up by $a_0(\tau)$. A similar effect has been discussed in \cite{Witten:1998zw} in the AdS/CFT context, where the conformal invariance guarantees to scale up the area of the contour without changing the world-sheet area in the bulk. 

Because the functions $f(r)\equiv f(r,\tau)/a_0(\tau)$ and $g(r)\equiv g(r,\tau)$ made out of the metric components diverge as $r\to\infty$, the action (\ref{action0}) also diverges with $\int^{\infty}{g(r)dr}$. We may relate this fact to the quark mass, that we should subtract from the action (\ref{action0}) \cite{Maldacena:1998im,Sonnenschein:2000qm}. So we define the regularized quark mass as 
\ben 
m_q=\frac{1}{2\pi \alpha'}\int_{0}^{r_1}{g(r)dr}=\frac{1}{2\pi \alpha'}\left[\int_{0}^{r_0}{g(r)dr}+\int_{r_0}^{r_1}{g(r)dr}\right],
\een
 such that the regularized action $S^{reg}=S-\frac{\sqrt{e}}{2\pi \alpha'}\int{m_q\,d\tau}$ is written as
\ben
 S^{reg}&=& \frac{\sqrt{e}T}{2\pi \alpha'}\left[\int_{r_0}^{r_1} dr\left(\frac{g(r)f(r)}{\sqrt{f(r)^2-f(r_0)^2}}-g(r)\right)-\int_{0}^{r_0}{g(r)dr}\right],
 \een 
where we set $r_1\to\infty$. Recall that from equation (\ref{fg}) the function $g$ is $\tau$ independent, so does the quark mass $m_q$.

%Therefore one may write also hereb$S=\sqrt{e}T E$.

%%%%%%%%%%%%%%%%%%%%%%%%%%%%%%%%%%%%%%%%%%%%%%%%%%%
\section{Large $N$ limit (or Large $D$-Brane Tension) }
%%%%%%%%%%%%%%%%%%%%%%%%%%%%%%%%%%%%%%%%%%%%%%%%%%%

Since we are interested in the weak version of the gauge/gravity duality \cite{D'Hoker:2002aw,Edelstein:2006kw}, we take into account only
the leading string contributions, where gravity description is valid. In this limit we have that the string 
coupling  $g_s\to0$ and the number of $D$-branes $N\to\infty$, while $g_sN$ is held constant and sufficiently large. As a consequence 
the $D3$-brane tension $T_{D3}=(2\pi)^{-3}l_s^{-4}g_s^{-1}$ should be very large. Notice that we try to maintain almost all the ingredients of the AdS/CFT conjecture, such that at least the holographic principle of gauge theory (on the boundary) corresponding to gravity (on the bulk) is guaranteed. 

Hereafter we assume that $T_{D3}=\sigma$. Since $\mu\sim\sqrt{\kappa^4\sigma^2}$, we have that $\mu$ is sufficiently large. This simplifies our computations by considering $\mu r$ also very large for $r>0$. Thus, for\footnote{There is another branch \cite{Bazeia:2007vx}, $r\leq 0$,
that we disregard in the present investigations.} $r> 0$, we may rewrite the warp factor as
\begin{equation}
U(r)=({e\gamma+\xi^2e^{\mu r}+\chi^2e^{-\mu r}})^{\frac{1}{2}},
\end{equation}
where
\begin{eqnarray}
	\label{xi-chi}
	\xi^2= \frac{1}{2}({1-e\gamma-\sqrt{1-2e\gamma}}\,), \qquad
	\chi^2= \frac{1}{2}(1-e\gamma+\sqrt{1-2e\gamma}\,).
\end{eqnarray}
Now, the derivative of the warp factor is given by
\begin{equation}
	U'=\frac{1}{R_0~U}\sqrt{\left(U^2 -e\gamma\right)^2-4\xi^2\chi^2},
\end{equation}
with a bulk characteristic radius defined as
\begin{equation}
	R_0=\frac{2}{\mu}.
\end{equation}
Thus, we have the helpful relation
\begin{equation}
	\label{diff-r-diff-U}
	dr=\frac{1}{U'}~dU=\frac{R_0~U}{\sqrt{\left(U^2 -e\gamma\right)^2-4\xi^2\chi^2}}~dU.
\end{equation}
It is useful to rescale $U$ to have dimension of energy by rearranging the dimensions of the quantities in the metric through the following transformations
\begin{eqnarray}
	\label{0}
	U&\to& \frac{\alpha'}{R_0}~U,\nonumber \\
	R_0^2 &=& \alpha' ~ R^2,
\end{eqnarray}
where $\alpha'$ is related to the inverse of the string tension. 
We now consider the following changing of coordinate
\begin{equation}\label{1}
	y=\frac{U}{U_0},\ \ \ \ \textrm{with } U_0\equiv U(r_0),
\end{equation}
with $r_0$ being where $U(r)$ assumes its minimum value. 

In terms of the above transformations, i.e., (\ref{diff-r-diff-U}), (\ref{0}) and (\ref{1}) the metric reads\footnote{According to transformations (\ref{0}), the dimensions of the quantities in this metric are the following: $[\alpha']=(\textrm{length})^2$, $[U]=(\textrm{length})^{-1}$, $[R]=1$, $[\tau]=[\vec{x}]=\textrm{length}$.}
\begin{equation}\label{def_metric}
	ds^2=\alpha'\left[\frac{U_0^2}{R^2} y^2 \left( e~dt^2+a_0(t)^2 d\vec{x}^2 \right) + \frac{R^2 y^2}{\left(y^2-C_+ \right)\left(y^2-C_- \right)}~dy^2\right],
\end{equation}
with
\begin{equation}
	\label{C+-}
	C_\pm=\left(e\gamma \pm 2|\xi \chi| \right)\frac{R^2}{\alpha' U_0^2}, 
\end{equation}
where $\xi$ and $\chi$ is given in (\ref{xi-chi}). Notice that as $\gamma\to0$ equations (4.2) and (4.9) tell us that the `deforming constants' $C_{\pm}\to0$. Furthermore equations (3.4) and (3.5) say that in this regime $a_0(t)$ is just a constant we set equal to 1. As a consequence the AdS space (2.2) is recovered from (4.8). As we see later, considering $\gamma$ small we find the interquark Cornell potential perturbativelly in $C_{\pm}$. The leading term is the same found in the original AdS/CFT correspondence. The limit $y\to\infty$ recovers the boundary of the five-dimensional space as the four-di\-men\-sion\-al Euclidean $AdS_4$ space (for $e=+1$) that we may identify as the induced metric on the 3-brane. 

%The Wilson loop is then evaluated on this space, which is associated with a $U(1)$ gauge theory and is far from the stack of $N$ coincident $D3$-branes at $y=1$ (associated with the delta function source) related to a $U(N)$ gauge theory.

%%%%%%%%%%%%%%%%%%%%%%%%%%%%%%%%%%%%%%%%%%%%%%%%%%%%%%%%%%%%%%%%%%%%%%%%%%%%%%%%%%%%%%%%%%%%%%%%%%%%%%%%%

We now proceed to evaluate the interquark distance and the action.

By using (\ref{stat_len}) and the transformations (\ref{diff-r-diff-U}), (\ref{0}) and (\ref{1}), we find that the static interquark distance is
\begin{equation}
	L_0=2\frac{R^2}{U_0}\int_1^{U_1/U_0} dy~\frac{1}{\sqrt{y^4-1}}~\frac{1}{\sqrt{\left(y^2-C_+ \right)\left(y^2-C_- \right)}}.
\end{equation}

By using (\ref{action0}) and the transformations (\ref{diff-r-diff-U}), (\ref{0}) and (\ref{1}), we find that the action is
\begin{equation}
	S=\frac{\sqrt{e}T~U_0}{2\pi}\int_1^{U_1/U_0} dy~ \frac{y^4}{\sqrt{y^4-1}}~\frac{1}{\sqrt{\left(y^2-C_+ \right)\left(y^2-C_- \right)}}.
\end{equation}

%The constants $C_\pm$ play an important role here. We   discuss different values for them that lead to different physics. 

As a first case, we can easily see that that the limits $C_\pm\to 0$ into the metric (\ref{def_metric}) recover the $AdS_5$ metric and the limit $y\to \infty$ recovers the $AdS_5$ boundary.

Since
\begin{equation}
	\xi^2\chi^2=\frac{1}{4}\left( (e\gamma)^2+1-2e\gamma-|1-2e\gamma| \right),
\end{equation}
which appears in the definition of $C_\pm$ in (\ref{C+-}), we may consider the following two cases:
\begin{enumerate}
	\item If $e\gamma\leq \frac{1}{2}$, then
	\begin{equation}
		\xi^2\chi^2=\frac{(e\gamma)^2}{4},
	\end{equation}
	and 
	\begin{equation}
		C_\pm=\left(e \pm 1 \right)~\gamma~\frac{R^2}{\alpha'U_0^2}.
	\end{equation}

	\item If $e\gamma>\frac{1}{2}$, then
	\begin{equation}
		\xi^2\chi^2=\frac{1}{4}\left(\gamma^2+2(1-2e\gamma) \right),
	\end{equation}
	and
	\begin{equation}
		C_\pm=\left(e\gamma\pm\sqrt{(e\gamma-2)^2-2} \right)\frac{R^2}{\alpha'U_0^2}
	\end{equation}

\end{enumerate}

%%%%%%%%%%%%%%%%%%%%%%%%%%%%%%%%%%%%%%%%%%%%
\section{Results}

In this section we analyse the two possible regimes for $e\gamma$ as stated in the end of the last section. From now on we mainly work in the Euclidean case, $e=+1$. In this case we show how a deconfining phase and a confining phase appear. We also evaluate the Cornell potential. As a last result we comment on the Lorentzian case, $e=-1$, and conjecture on interesting possible phenomenology for meson pair production. 

Following the common practice initiated in \cite{Maldacena:1998im}, we evaluate both integrals for the action and interquark distance in the Euclidean space. Thus, we have for 
\begin{enumerate}
	\item $\gamma\leq \frac{1}{2}$ that
	\begin{eqnarray}
		C_+&=& 2\gamma~\frac{R^2}{\alpha'U_0^2}, \\
		C_-&=& 0,
	\end{eqnarray} 
	and for
	\item $\gamma>\frac{1}{2}$ that
	\begin{equation}
		C_\pm=\left(\gamma\pm\sqrt{(\gamma-2)^2-2} \right)\frac{R^2}{\alpha'U_0^2}= \left(1\pm\sqrt{(1-\frac{2}{\gamma})^2-\frac{2}{\gamma^2}} \right)~\gamma~\frac{R^2}{\alpha'U_0^2}.
	\end{equation}
	Here we may consider $\gamma\to\infty$, so that
	\begin{equation}
		C_\pm=\left(1\pm 1 \right)~\gamma~\frac{R^2}{\alpha'U_0^2},
	\end{equation}
which simplifies to
	\begin{eqnarray}
		C_+&=&2\gamma \frac{R^2}{\alpha'U_0^2} \\
		C_-&=& 0.
	\end{eqnarray}
\end{enumerate}
In summary, for any of the two cases above, the integrals we have to study simply become
\begin{eqnarray}
	\label{act}
	L_0&=&2\frac{R^2}{U_0}\int_1^{U_1/U_0} dy~\frac{1}{y\sqrt{y^4-1}}~\frac{1}{\sqrt{y^2-C_+}}, \\
	S&=&\frac{T~U_0}{2\pi}\int_1^{U_1/U_0} dy~ \frac{y^3}{\sqrt{y^4-1}}~\frac{1}{\sqrt{y^2-C_+}}\label{len}.
\end{eqnarray}
In  the sequel, we shall evaluate the interquark potential using the above integrals. First we obtain a Coulomb-like phase (deconfining phase). Next we obtain a confining phase.

\subsection{A Coulomb-like Phase}

From now on, we consider
\begin{equation}
\label{eq0}
\qquad \gamma\leq1/2,\qquad C_+=2\gamma\frac{R^2}{\alpha'U_0^2}, 
\end{equation}
so that the interquark distance is now
\begin{equation}
\label{glesshalf1} 
L_0=2\frac{R^2}{U_0}\int_1^{U_1/U_0}{dy\frac{1}{y\sqrt{y^4-1}}\frac{1}{\sqrt{y^2-C_+}}}~~\stackrel{U_1\to\infty}{=}~~2\frac{R^2}{U_0}I_1(C_+).
\end{equation}
The regularized action is now
\begin{equation}
\label{glesshalf2} 
S^{reg}=\frac{T~U_0}{2\pi}\left[\int_1^{U_1/U_0}{dy\left(\frac{y^3}{\sqrt{y^4-1}}\frac{1}{\sqrt{y^2-C_+}}-1\right)}-1\right]~~\stackrel{U_1\to\infty}{=}~~\frac{T~U_0}{2\pi}I_2(C_+),
\end{equation}
where $I_{1,2}(C_+)$ being definite integrals for $U_1\to\infty$.

Now, for $C_+=0$ (i.e., $\gamma=0$), we solve (\ref{glesshalf1}) for $U_0$, and substitute it into (\ref{glesshalf2}) in order to find
\begin{equation}
\label{eq3} 
S^{reg}\equiv TE(L_0)=\frac{T}{\pi}~I_1(0)I_2(0)~\frac{R^2}{L_0}\longrightarrow E(L_0)=\,-{\frac {{4\pi }^{2}}{{\Gamma} \left( 1/4 \right)^{4}}\frac{R^2}{L_0}}.
\end{equation}
Thus, the regularized action (\ref{glesshalf2}) gives the interquark potential in a Coulomb-like phase. This case gives precisely the 
Maldacena's results for AdS/CFT correspondence \cite{Maldacena:1998im}, with $R_0^4\equiv R^4~\alpha'^2=4\pi g_s N~\alpha'^2=4\pi g^2_{YM} N~\alpha'^2$, with $g^2_{YM}=g_s$. Recall that the Maldacena limit is defined as $R$ large and fixed, when $N\to\infty$, while $\alpha'\to 0$. This is consistent with $R_0\to 0$, since either $R_0=2/\mu$ and $\mu\to\infty$ or $R_0^2=\alpha'R^2$ and $\alpha\to 0$ with $R$ large and fixed.

\subsection{Opening Up a Confining Phase}

Let us now consider that $C_+=1$, i.e., $U_0^2=2\gamma \frac{R^2}{\alpha'}$. %This, of course, forces $R_0=U_0$. 
This case reveals a new regime in the calculations above. Firstly we note that the integral (\ref{act}) becomes now
\begin{equation}
\label{eq4} 
L_0=\frac{\alpha'U_0}{\gamma}\int_1^{U_1/U_0}{dy\frac{1}{y\sqrt{y^4-1}}\frac{1}{\sqrt{y^2-1}}}~~\stackrel{U_1\to\infty}{=}~~\frac{\alpha'U_0}{\gamma}I_1(1),
\end{equation}
(note that the prefactor of $L_0$ has changed since $2R_0^2/U_0\to \frac{\alpha'U_0}{\gamma}$), and the regularized action is now
\begin{equation}
\label{eq5} 
S^{reg}=\frac{T~U_0}{2\pi}\left[\int_1^{U_1/U_0}{dy\left(\frac{y^3}{\sqrt{y^4-1}}\frac{1}{\sqrt{y^2-1}}-1\right)}-1\right]~~\stackrel{U_1\to\infty}{=}~~\frac{T~U_0}{2\pi}I_2(1),
\end{equation}
where $I_{1,2}(C_+=1)$ being definite integrals for $U_1\to\infty$. Just as in the previous calculation, we can easily show that the regularized action (\ref{eq5}) written in terms of the length $L_0$ (\ref{eq4}) is
\begin{equation}
\label{eq6} 
S^{reg}={T}\frac{I_2(1)}{I_1(1)}~\frac{\gamma}{2\pi \alpha'}~L_0.
\end{equation}
The integrals $I_{1,2}(1)$ are regularized only in the UV-regime. There is a pole at $y=1$ that should also be regularized. This can easily be done 
by considering an infrared cut-off $\varepsilon$ that can be set equal to `1' in the end of the computations (see Appendix). We anticipate that this procedure will
reveal that the rate $\frac{I_2(1)}{I_1(1)}\to1$ as $\varepsilon\to1$. Thus, we end up with the confining regime
\begin{equation}
\label{eq7} 
S^{reg}\equiv E(L_0)T={T}\frac{\gamma}{2\pi \alpha'}~L_0\longrightarrow E(L_0)=\frac{\gamma}{2\pi \alpha'}~ L_0,
\end{equation} 
where $\gamma/2\pi \alpha'$ plays the role of a ``QCD string" tension, that depends on the $D3$-brane tension and the Hubble parameter.
We note that this result can also be written as 
\begin{equation}
\label{Econf} 
E(L_0)=\frac{\gamma~a_0(\tau)}{2\pi \alpha'}~L(\tau).
\end{equation}
where we have used $L_0=a(\tau)L(\tau)$. This generalizes the confining static term $E(L_0)=f(r_0)L_0$ obtained
in the literature. In the present case, one still have a static confining term that depends on the cosmological constant $\bar{\Lambda}=H_E^2\sim\gamma \sigma$ on the $D3$-brane
where the field theory lives.

\subsection{Cornell Potential}
%%%%%%%%%%%%%%%%%%%%%%%%%%%%%%%%%%%%%%%%

Let us now consider the equation (\ref{act}) and (\ref{len}) to evaluate the Cornell potential. Here we expand these integrals in powers of $C_+$
with
\begin{equation}
\label{Eq0}
\qquad \gamma\leq1/2,\qquad C_+=2\gamma\frac{R^2}{\alpha'U_0^2}. 
\end{equation}
For the sake of simplicity we are just considering the leading term in the interquark distance $L_0$ i.e.
\begin{equation}
\label{Eq1} 
L_0=2\frac{R^2}{U_0}\int_1^{U_1/U_0}{dy\frac{1}{y^2\sqrt{y^4-1}}}+O(C_+)\stackrel{U_1\to\infty}{=}~~2\frac{R^2}{U_0}I_1(0)+O(C_+),
\end{equation}
and regularized action power expanded up to the second order in $C_+$
\begin{eqnarray}
\label{eq2} 
S^{reg}=\frac{T~U_0}{2\pi}\left[\int_1^{U_1/U_0}{dy\left(\frac{y^2}{\sqrt{y^4-1}}-1\right)}-1+\frac{C_+}{2}\int_1^{U_1/U_0}{dy\frac{1}{\sqrt{y^4-1}}}\right.\nonumber\\
+\left.\frac{3C_+^2}{8}\int_1^{U_1/U_0}{dy\frac{1}{y^2\sqrt{y^4-1}}}\right]+O(C_+^3).
\end{eqnarray}
Thus,
\begin{equation}
	E(L_0)=\frac{I_1 I_2}{2\pi} \frac{R^2}{L_0}+\frac{I_3}{2I_1} \frac{\gamma~L_0}{2\pi \alpha'}+\frac{3}{16} \frac{1}{I_1^2} \frac{\gamma^2~L_0^3}{2\pi \alpha'^2~R^2},
\end{equation}
where
\begin{eqnarray}
	I_1&=&\int_1^{U_1/U_0}{dy\frac{1}{y^2\sqrt{y^4-1}}}=1/2\,{\frac {\Gamma  \left( 3/4 \right) ^{2}\sqrt {2}
}{\sqrt {\pi }}}\stackrel{U_1\to\infty}{=}~~0.599070117, \nonumber\\
	I_2&=&\int_1^{U_1/U_0}{dy\left(\frac{y^2}{\sqrt{y^4-1}}-1\right)}-1\stackrel{U_1\to\infty}{=}~~-1/2\,{\frac { \Gamma  \left( 3/4 \right) ^{2}\sqrt {2
}}{\sqrt {\pi }}}=-I_1, \nonumber \\
	I_3&=&\int_1^{U_1/U_0}{dy\frac{1}{\sqrt{y^4-1}}} = 1/4\,{\frac {{\pi }^{3/2}\sqrt {2}}{ \Gamma  \left( 3/4
 \right) ^{2}}}\stackrel{U_1\to\infty}{=}~~1.311028777. \nonumber
\end{eqnarray}

\begin{comment}
Thus, \look 
\ben
\label{cornell}
E(L_0)=\frac{1}{2\pi\alpha'}\left[- 1.198140233~\frac { \alpha'R^2}{L_0}+ 1.094219807~L_0\gamma +
 0.5224514746~\frac{L_0^3\gamma^2}{\alpha'R^2}\right]
\een
\end{comment}
Thus, 
\ben
\label{cornell}
E(L_0)=\frac{1}{2\pi\alpha'}\left[- 0.358885005~\frac { \alpha'R^2}{L_0}+ 1.094219807~L_0\gamma +
 0.5224514746~\frac{L_0^3\gamma^2}{\alpha'R^2}\right]
\een

This is the well-known \cite{Eichten:1974af} Cornell potential $E(r)=-a/r + b r,$
where $a$, $b$ are nonnegative constants and $r$ is the interquark distance.
This potential is commonly used to describe the physics of heavy
quarks. One can successfully obtain the whole mass spectrum of the quark
anti-quark pair in the quarkonium system using this potential. In the confining term in (\ref{cornell}),
there appears $\gamma$ that depends on the Euclidean Hubble parameter
$H_E$ and the brane tension $\sigma$. Thus `QCD string' tension is $\sim \gamma/\alpha'$.

\begin{comment}
Assuming a quarkonium system one can estimate the suitable regime for the
cosmological constant favoring the
meson pair production
during the evolution of the Universe as $\gamma/\alpha' L_0\to \textrm{const.}$,
being $L_0$ the typical meson size in this
regime. Several other investigations can also be done such as relating the
number of e-folds to the confining phase.
\end{comment}

%%%%%%%%%%%%%%%%%%%%%%%%%%%%%%%%%%%%%%%%%%%%%%%%%%%%%%%%%%%%%%%%%%%%%%
\subsection{A Lorentzian signature limit}
%%%%%%%%%%%%%%%%%%%%%%%%%%%%%%%%%%%%%%%%%%%%%%%%%%%%%%%%%%%%%%%%%%%%%%

There is another interesting limit. In the limit of $C_\pm\to\infty$ ($\gamma\to\infty$) the Nambu-Goto action is well defined only in a `Lorentzian space', i.e. $e=-1$, where $C_+\to C_-$. Thus, the leading term of the properly regularized action (\ref{eq2}) in the limit $C_-\to\infty$ reads 
\begin{equation}
\label{eq2.1} 
S^{reg}=\frac{\sqrt{e}}{\sqrt{-C_-}}\frac{T~U_0}{2\pi}\int_1^{U_1/U_0}{dy\left(\frac{y^3}{\sqrt{y^4-1}}-y\right)}=\frac{\sqrt{e}}{2\sqrt{-C_-}}\frac{TU_0}{2\pi}=
\frac{\sqrt{\alpha'}}{2\sqrt{2\gamma}}~\frac{iTU_0^2}{2\pi R},%=\frac{TU_0^2}{2\sqrt{2|\gamma|} R_0},
\end{equation}
where we have used the previous result $C_\pm=(e\pm1)\frac{\gamma R^2}{\alpha'U_0^2}$ and the fact that $\sqrt{e}=i$ %and $\gamma=-|\gamma|$ 
for the Lorentzian signature. 
%Recall that we have early assumed our brane with negative cosmological constant such that $\bar{\Lambda}=H_E^2=-H^2\sim \gamma\sigma$. 
Thus the mass of a meson pair is obtained by using the Wilson loop evaluated in the Lorentzian signature \ben <W({\cal C})>\sim e^{i\,TE(L)}=e^{S^{reg}}, \een such that
\ben \label{mes}
E(L)=\frac{\sqrt{\alpha'}}{\sqrt{2|\gamma|}}~\frac{U_0^2}{4\pi R}.
\een
Recall that $U_0$ assumes its minimum value at $r_0$ in the $AdS_5$ space. We can see that in the limit of $C_\pm\to\infty$ in (\ref{act}) $U_0$ does not depend on $L_0$. Thus, we can ensure that the potential in (\ref{mes}) does not depend on the interquark distance $L_0$ and is therefore constant. This regime is usually assumed
to be favorable to meson pairs creation from the vacuum \cite{Erdmenger:2007cm,Karch:2002xe}.
This is a regime where the Wilson loop develops a `perimeter law', in contrast to the area law in the
confining regime.

%%%%%%%%%%%%%%%%%%%%%%%%%%%%%%%%%%%%%%%%%%%%%%%%%%
\section{Conclusions and Perspectives}
%%%%%%%%%%%%%%%%%%%%%%%%%%%%%%%%%%%%%%%%%%%%%%%%%%%%

In this paper we have considered the five-dimensional space, the brane
cosmology background, as a gravity dual of a confining gauge field theory living on its
boundary. This space develops the precise phenomenon of deforming a non-confining conformal AdS
geometry into a confining geometry described by an asymptotically Euclidean $AdS_4$ -
$AdS_5$ space. The main result shows that the cosmological constant on the brane is directly
related to the confining term of the interquark potential. 

The well-known Cornell potential found in earlier proposals for AdS/QCD correspondence \cite{Andreev:2006ct} was also here obtained.  It is the kind of potential we can use to describe heavy quarks. One can successfully obtain the whole mass spectrum of the quark anti-quark pair in the quarkonium system using such a potential. Thus, in our setup it is possible to estimate the suitable regime for the
cosmological constant favoring the meson pair production
during the evolution of the Universe as $(\gamma/\alpha')L_0\to \textrm{const.}$,
with $L_0$ being the typical meson size in this regime. Several other investigations can also be done such as relating the
number of e-folds with the confining phase.

We have also obtained a regime where the Wilson loops shows up a `perimeter law'. As pointed out in \cite{Erdmenger:2007cm,Karch:2002xe}, it is possible to consider meson pair creation in this regime. Furthermore, as one can see in (\ref{mes}), this regime also depends on the Hubble parameter.

In summary, in our setup we show an expected phenomenology described in the dual field theory for the Universe in the inflationary brane. As one knows, in the inflationary regime, the Universe is cooling. Therefore there is a phase transition in QCD - from a phase of nearly free quarks and gluons to a hadronic phase. Thus, it is tempting to imagine that as the Universe expands quarks may confine, in some delicate manner, because of the accelerated expansion of the Universe due to a cosmological constant.
%\acknowledgments
%\noindent
\subsection*{Acknowledgments}
The work of F.A.B. is supported in part by CNPq and PROCAD/PNPD-CAPES. A.R.Q. would like to thanks Prof. Bruno Carneiro Cunha for useful discussions. A.R.Q would also like to thank the staff at the ICCMP. LB and FAB thanks ICTP for financial support and hospitality during the early stages of this work.
%%%%%%%%%%%%%%%%%%%%%%%%%%%%%%%%%%%%%%%%%%%%%%%%%%%%%%%%%%%%%%%%%
\section{Appendix: Evaluating the integrals $I_{1,2}(1)$}
%%%%%%%%%%%%%%%%%%%%%%%%%%%%%%%%%%%%%%%%%%%%%%%%%%%%%%%%%%%%%%%%%

Let us use the infrared cut-off $\varepsilon$ in order to properly evaluate the integrals $I_{1,2}(1)$.
The first integral is
\begin{eqnarray}
\label{eq8} 
I_1(1)&=&\int_\varepsilon^{\infty}{dy\frac{1}{y\sqrt{y^4-1}}\frac{1}{\sqrt{y^2-1}}}=1/4\,\sqrt {2}{ \rm arctanh}\left( 1/2\,{\frac {\sqrt {2}\varepsilon}{
\sqrt {{\varepsilon}^{2}+1}}}+1/2\,{\frac {\sqrt {2}}{\sqrt {{\varepsilon}^{2}
+1}}} \right)\nonumber\\
&-&1/4\,\sqrt {2}{\rm arctanh} \left( 1/2\,{\frac {\sqrt {
2}\varepsilon}{\sqrt {{\varepsilon}^{2}+1}}}-1/2\,{\frac {\sqrt {2}}{\sqrt {{
\varepsilon}^{2}+1}}} \right) -{\rm arctanh} \left( {\frac {1}{\sqrt {{
\varepsilon}^{2}+1}}} \right) 
,
\end{eqnarray}
and
\begin{eqnarray}
\label{eq9} 
I_2(1)&=&\int_\varepsilon^{\infty}{dy\left(\frac{y^3}{\sqrt{y^4-1}}\frac{1}{\sqrt{y^2-1}}-1\right)}-1=-{\frac {{\varepsilon}^{2}}{\sqrt {{\varepsilon}^{2}+1}}}-{\frac {1}{\sqrt {{
\varepsilon}^{2}+1}}}\nonumber\\
&+&1/4\,\sqrt {2}{\rm arctanh} \left( 1/2\,{\frac {
\sqrt {2}\varepsilon}{\sqrt {{\varepsilon}^{2}+1}}}+1/2\,{\frac {\sqrt {2}}{
\sqrt {{\varepsilon}^{2}+1}}} \right) \nonumber\\
&-&1/4\,\sqrt {2}{\rm arctanh} \left( 
1/2\,{\frac {\sqrt {2}\varepsilon}{\sqrt {{\varepsilon}^{2}+1}}}-1/2\,{\frac {
\sqrt {2}}{\sqrt {{\varepsilon}^{2}+1}}} \right) +\varepsilon-1.
\end{eqnarray}
In the limit $\varepsilon\to1$ both integral goes to infinity with the same dominant term: 
\begin{equation}
{ \rm arctanh}\left( 1/2\,{\frac {\sqrt {2}\varepsilon}{
\sqrt {{\varepsilon}^{2}+1}}}+1/2\,{\frac {\sqrt {2}}{\sqrt {{\varepsilon}^{2}
+1}}} \right)\to\infty.
\end{equation}
Thus, recovering the original limit of the integral one finds $\frac{I_1(1)}{I_2(1)}\to 1$.

%\bibliographystyle{JHEP}
%\bibliography{BBQ08-ADS-QCD}
\providecommand{\href}[2]{#2}\begingroup\raggedright\endgroup

\end{document}